\documentclass{osa-article}

\journal{osajournal}
\articletype{Research Article}
\usepackage{lineno}
\usepackage{caption}
\usepackage{subcaption}
\usepackage{siunitx}

\begin{document}

\title{Analog Stabilization of an Electro-Optic I/Q Modulator with an Auxiliary Modulation Tone}

\author{Sebastian Wald,\authormark{1,*} Fritz Diorico,\authormark{1} and Onur Hosten\authormark{1}}

\address{\authormark{1}Institute of Science and Technology Austria (ISTA), Am Campus 1, 3400 Klosterneuburg, Austria\\

}

\email{\authormark{*}sebastian.wald@ista.ac.at} %% email address is required



\begin{abstract*}
Proper operation of electro-optic I/Q modulators rely on precise adjustment and control of the relative phase biases between the modulator's internal interferometer arms. 
We present an all-analog phase bias locking scheme where error signals are obtained from the beat between the optical carrier and optical tones generated by an auxiliary \SI{2}{MHz} \textit{RF}-tone to lock the phases of all three involved interferometers for operation up to \SI{10}{GHz}.
With the developed method, we demonstrate an I/Q modulator in carrier-suppressed single-sideband mode, where the suppressed carrier and sideband are locked at optical power levels $< -27 dB$ relative to the transmitted sideband.
We describe a simple analytical model for calculating the error signals, and detail the implementation of the electronic circuitry for the implementation of the method. 

\end{abstract*}

%%%%%%%%%%%%%%%%%%%%%%%%%%  body  %%%%%%%%%%%%%%%%%%%%%%%%%%
\section{Introduction}
Electro-optic modulators (EOMs) are key elements for modern telecommunication technologies \cite{Wooten2000}.
In atomic, molecular, and optical physics, they are widely used to lock lasers to optical cavities \cite{Drever1983}, to frequency offset lock lasers \cite{Johnson2010, Greve2022} and to shift the frequency of a single laser \cite{Zhu2018,Templier2021}. 
%Common EOMs are fiber-coupled waveguides integrated in an optically non-linear crystal, e.g. lithium niobate, as substrate.
%Since the optical refractive index of such non-linear elements can be tuned by an applied electric field, EOMs enable 
Waveguide based EOMs integrating nonlinear crystals, such as lithium niobate, enable broadband electronic phase modulation of the transmitted light up to tens of \SI{}{\giga\hertz} \cite{Buhrer1962}. 
More complex waveguide structures allow further applications. 
A single Mach-Zehnder interferometer (MZI) structure enables amplitude modulation and a dual-parallel MZI structure gives simultaneous access to the phase and amplitude of the transmitted light (I/Q modulator) \cite{Izutsu1981, Cusack2004}. 
The phase bias between the internal MZI arms define the operating mode of the modulator. For example, an I/Q modulator can be operated in either single-sideband (SSB) or carrier-suppressed single-sideband mode (CS-SSB).       

Internal charge and external environmental fluctuations can induce drifts of these phase biases, that in turn, reduce the effectiveness of amplitude and I/Q modulators \cite{Svarny2010}. 
Especially for high-precision experiments, these drifts can induce phase and amplitude noise. 
Hence, active stabilization of the phase biases is required. 
Prior and commercially available stabilization schemes rely on weak modulation of the phase bias voltage, adding dither tones in the \SI{}{\kilo\hertz} range with an amplitude of several tens or hundreds of \SI{}{mV} to the bias voltage. 
This modulation is transferred to the optical carrier and an error signal can be extracted by the monitored optical signal. 
In \cite{Fabbri2013}, bias voltages were modulated with different dither frequencies and stabilized with lock-in amplifiers.
Other commercially available feedback modules (Thorlabs, MX10A; iXBlue, MBC-IQ-LAB) modulate the phase biases with a single frequency dither and obtain the error signals by digital analysis of the optical spectrum \cite{Bui2011}.

In this paper, we make use of a small auxiliary modulation to realize a stabilization method for electro-optic amplitude and I/Q modulators instead of dithering the phases of the device's interferometers.
In particular, using a telecom band electro-optic I/Q modulator, we apply an auxiliary \SI{2}{MHz}-modulation to generate a set of sidebands (SB) in addition to the primary modulation sidebands of interest that lie in the GHz range. 
The beat signals between these sidebands and the optical carrier allow us to generate the phase bias error signals using standard \textit{RF}-mixers.
A home built, all-analog feedback circuit receives the error signals and stabilizes the corresponding phase biases.
The single auxiliary tone allows us to efficiently stabilize all three interferometers of the modulator in comparison to other methods, and in principle permits smaller modulations due to bypassing technical noise problems at \SI{}{kHz} frequencies.

Below, we first review the use of an I/Q  modulator as a CS-SSB generator, and describe our method of obtaining error signals for all three phase biases. 
Then we discuss the experimental set up and the developed circuitry including the home built demodulation and feedback circuits. 
Finally, we show experimental results for the generated error signals and demonstrate the stability of the locked modulator in CS-SSB mode.

\section{Materials and Methods}

\subsection{Electro-optic I/Q modulator}

\begin{figure}[b]
    \centering
    \includegraphics[width=\linewidth]{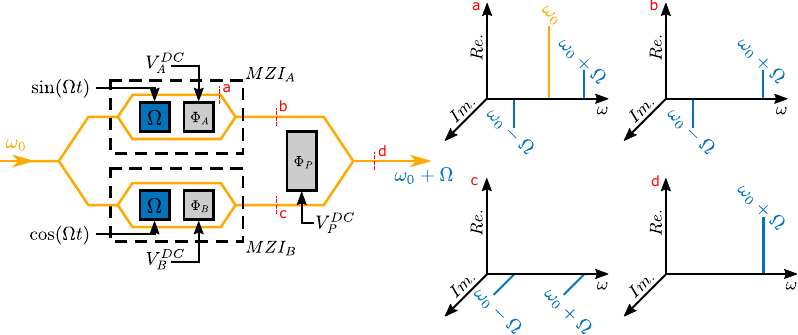}
    \caption{Left: Schematic of an electro-optic I/Q modulator. 
    The dual-parallel MZI structure enables simultaneous amplitude and phase modulation. 
    For CS-SSB generation, the sub-MZIs ($MZI_{A,B}$) are modulated with frequency $\Omega$ and a relative phase of $\pi / 2$. 
    By tuning the control voltages $V^{DC}_{A,B,P}$, the corresponding phase biases are adjusted to $\Phi_{A,B} = \pi$ and $\Phi_P = \pi/2$. 
    The optical carrier ($\omega_0$) and one sideband ($\omega_0-\Omega$) are suppressed. 
    Right: Optical frequency spectra at the marked locations in the schematic.
    (a) Phase modulation of the carrier.
    (b,c) Output of $MZI_{A,B}$. The phase bias $\Phi_{A,B} = \pi$ bewteen their arms leads to destructive interference of the carrier. 
    (d) The combination of the sub-MZI with a phase bias $\Phi_P = \pi/2$ generates a single-sideband.}
    \label{fig:1}
\end{figure}

The dual-parallel Mach-Zehnder structure of a fiber-coupled, electro-optic I/Q modulator is illustrated in Fig. \ref{fig:1}. 
Each sub-MZI ($MZI_{A,B}$) can be individually modulated via separate \textit{RF}-ports. 
The three \textit{DC}-ports ($V^{DC}_{A,B,P}$) allow tuning of the relative phase between the MZI arms ($\Phi_{A.B}$) and between the sub-MZIs ($\Phi_{P}$). 
By applying an \textit{RF}-tone with frequency $\Omega$, both arms of a single sub-MZI are phase modulated. 
For CS-SSB generation, the applied \textit{RF}-modulation has a $\pi/2$-phase between the two \textit{RF}-ports. 
This shifts the phase of the generated sidebands by 90 degrees between the two sub-MZIs as depicted in Fig. \ref{fig:1}.
The generated sidebands within each sub-MZI have a relative phase of $\pi$ between the upper and lower arms. 
This allows the optical carrier with frequency $\omega_0$ to destructively interfere at the outputs of the sub-MZIs while the sidebands interfere constructively when the phase bias between the arms is tuned to $\Phi_{A,B}=\pi$. 
With the adjustment of the phase bias between the arms of the outer-MZI to $\Phi_P = \pi/2$ or $\Phi_P = -\pi/2$, the upper or lower sideband destructively interferes and the output of the modulator is a CS-SSB with the frequency $\omega_0+\Omega$ or $\omega_0-\Omega$ respectively.

\subsection{Error signal generation}
Inadvertent drifts in the phase biases require their continuous adjustment over the course of hours.
To sense and correct for these drifts we apply a weak auxiliary modulation at a relatively low frequency ($\Omega_{LF} = 2 \pi \times \SI{2}{MHz}$) that generates sidebands, which beat with each other and the carrier. 
In order to describe this beat signal, we assume a single tone optical carrier at the I/Q modulator input, represented by the electric field amplitude $E_{in} = E_0 \exp{(-i \omega_0 t)}$. 
We evaluate the output for a single MZI, where the phase shifts for the upper and lower arms are $\pm \Phi/2$, where $\Phi = \phi +\pi +\delta$. 
Here, $\phi$ is the phase modulation, $\pi$ is the target phase bias between the MZI-arms and $\delta$ is a small perturbation of the phase bias.
The output field of such a MZI is given by
\begin{equation}
    E^{MZI}_{out} = \frac{E_{in}}{2} \big[ \exp(-i\Phi/2) +\exp(i\Phi/2) \big].
    \label{eq:MZI}
\end{equation}
In case of small $\phi$ and $\delta$, Eq. \ref{eq:MZI} can be simplified using the linear terms of its Taylor expansion.
The output fields of the two sub-MZIs of our I/Q modulator can then be expressed as:
\begin{equation}
    \begin{aligned}
        E_{out}^{MZI_{A}} = -\frac{E_0}{4}  \big[ 2 \beta \exp(-i \omega_0 t) \sin(\Omega_{LF} t)  +\delta_A   \exp(-i \omega_0 t) \big]        \\
        E_{out}^{MZI_{B}} = -\frac{E_0}{4} \big[ 2 \beta  \exp(-i \omega_0 t)  \cos(\Omega_{LF} t)  +\delta_B  \exp(-i \omega_0 t)  \big] 
    \end{aligned}
    \label{eq:MZI_out}
\end{equation}
Here, $\delta_{A}$ is the phase bias error of $MZI_A$, $\delta_{B}$ is the phase bias error of $MZI_B$, $\beta$ is the modulation depth, $\Omega_{LF}$ is the modulation frequency and $\sin(\Omega_{LF} t)$, $\cos(\Omega_{LF} t)$, respectively originates from the phase of the modulation signal. 
As seen from the last terms of Eqs. \ref{eq:MZI_out}, the residual optical carrier at the sub-MZI outputs are directly proportional to the phase bias errors.
The bias errors do not affect the sidebands (the first terms of Eqs. \ref{eq:MZI_out}) at this order of approximation.
Next, we calculate the output field of the complete I/Q modulator. 
The phase bias for the outer MZI is given by $\Phi_P = \pi/2 +\delta_P$. 
Again, we apply the first order approximation for the small phase bias error $\delta_P$ and we get the following expression for the output field:      
\begin{equation}
    E_{out}^{tot} = -\frac{E_0}{4} \exp[-i( \omega_0 t + \pi/4)]  \big[  2 \beta  \exp(i\Omega_{LF} t) - i \beta \delta_P  \exp(-i\Omega_{LF} t)  + i \delta_A +  \delta_B  \big]
    \label{eq:field_IQ_out}
\end{equation}

The power which is detected with a photodiode is proportional to $|E_{out}^{tot}|^2$. 
Retaining only first order terms in the phase errors, we get:
\begin{equation}
|E_{out}^{tot}|^2 = \frac{E_0^2}{4} \big[ \beta^2 + \delta_A \beta \sin(\Omega_{LF} t) +  \delta_B \beta \cos(\Omega_{LF} t) - \delta_P \beta^2 \sin(2\Omega_{LF} t) \big] 
    \label{eq:P_out}
\end{equation}
Eq. \ref{eq:P_out} shows that the sine and cosine quadratures of the beatnote observed at the photodiode at frequency $\Omega_{LF}$ give access to the phase bias errors $\delta_A$, $\delta_B$, and that the beatnote tone at frequency $2 \times \Omega_{LF}$ gives access to the phase bias error $\delta_P$.
Each of these bias errors can be extracted independently by demodulating three copies of the beatnote signal. 

In principle, it is possible to use the primary modulation for error signal generation.
However, the primary modulation often gets tuned or switched on and off during an experimental sequence. 
Thus, an additional weak auxiliary modulation can be applied to continuously generate an error signal for the phase biases. 
We choose the frequency of this modulation to be \SI{2}{MHz}, where it can be conveniently processed with standard \textit{RF}-components. 

\begin{figure}[t]
    \centering
    \includegraphics[width=\linewidth]{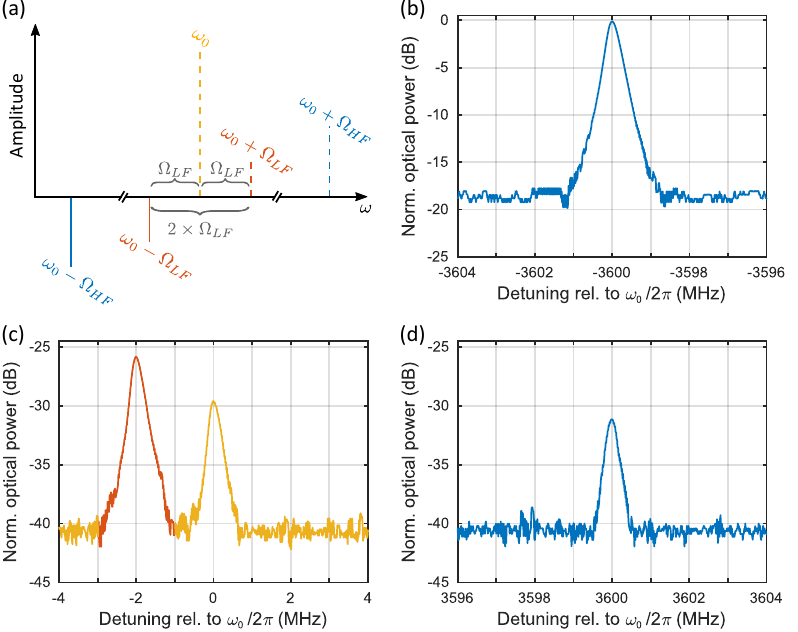}
    \caption{
    (a) Schematic frequency spectrum with frequencies $\Omega_{HF}$ (blue) and $\Omega_{LF}$ (red). 
    Dashed lines indicate suppressed frequency components.
    The upper and lower auxiliary SBs beat with the carrier (yellow) at frequency $\Omega_{LF}$. Choosing the correct demodulation phase gives access to the phase bias errors $\delta_{A}$ and $\delta_{B}$.
    The beat between the two auxiliary SBs at frequency $2 \Omega_{LF}$ is proportional to the phase bias error $\delta_P$.
    (b) Optical spectrum for the maximized lower primary sideband ($\Omega_{HF}$-SB) in the CS-SSB configuration.
    (c) Optical spectrum of the carrier including the auxiliary SBs. 
    The carrier is suppressed by \SI{-29.3}{dB} relative to the maximized SB. 
    Optimization of the lower $\Omega_{HF}$-SB  maximizes the lower auxiliary $\Omega_{LF}$-SB (red) as well. 
    The upper auxiliary SB is suppressed below the noise floor in this measurement.   
    (d) Optical spectrum showing \SI{-30.7}{dB} suppression of the upper $\Omega_{HF}$-SB. 
    The different noise floors in (b-c) originate from varying the signal resolution of the oscilloscope used for measurement.}
    \label{fig:2}
\end{figure}

The schematic frequency spectrum in presence of both the auxiliary ($\Omega_{LF}$) and the primary (high frequency, $\Omega_{HF}$) modulations is shown in Fig. \ref{fig:2}a. 
The modulations are depicted in a configuration where the lower sidebands are transmitted, with the dashed lines indicating suppressed frequency components. 
Drifts of the phase biases cause reappearance of the suppressed components as dictated by Eq. \ref{eq:P_out}. 
However, utilization of the generated error signals in a feedback loop that controls the phase biases acts to keep these components suppressed -- by keeping the condition $\delta_{A,B,P}=0$. 
Importantly, due to the broadband nature of the modulator, the sidebands at  $\Omega_{LF}$  and  $\Omega_{HF}$ are always correlated: suppressing a sideband at $\Omega_{LF}$ also suppresses the corresponding sideband at $\Omega_{HF}$.
Furthermore, potential deviations from this ideal behavior can easily be corrected by fine tuning the locking points of the phase biases.

\subsection{Experimental setup}
\begin{figure}[b]
    \centering
    \includegraphics[width=\linewidth]{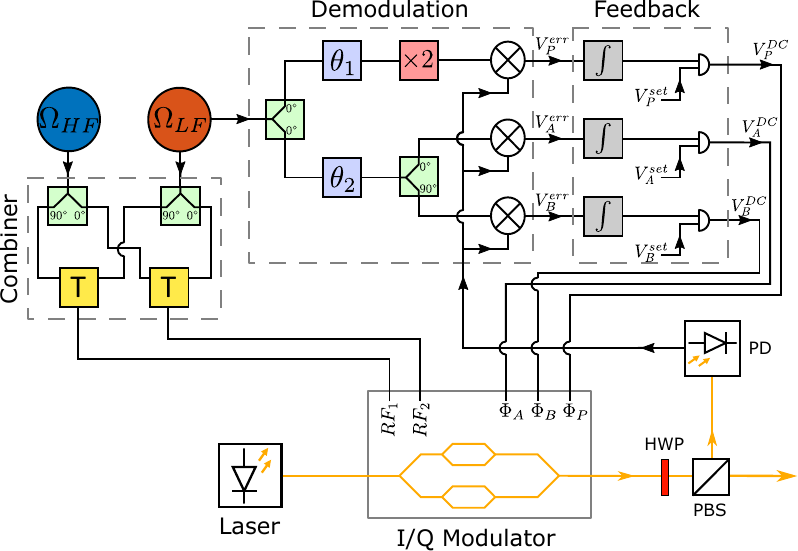}
    \caption{Schematic experimental setup for CS-SSB stabilization. 
    The I/Q modulator receives both the auxiliary ($\Omega_{LF}$) and the primary ($\Omega_{HF}$) modulations. 
    The optical beat signals of the $\Omega_{LF}$ sidebands are detected with a photodiode (PD). 
    The phase bias error signals ($V^{err}_{A,B,P}$) are generated with a demodulation circuit that generates the demodulation tones with adjustable phases ($\theta_1$, $\theta_2$) for cross talk suppression.
    The feedback circuit adjusts the phase bias control voltages ($V^{DC}_{A,B,P}$) and stabilizes the CS-SSB operation. 
    HWP, half-wave plate; PBS, polarizing beam splitter.
    Part list -- $\Omega_{LF}$-Splitters: Mini-circuits PSC-2-2+, PSCQ-2-8+ and PSC-3-2+; Phase shifters: Mini-circuits JCPHS-2.5+; Frequency doubler: Mini-circuits RK-3+; Mixers: Mini-circuits RPD-1+; $\Omega_{HF}$-Splitter:, Mini-circuits QCN-45D+; Bias tee (T): Mini-circuits TCBT-14+; Opamps in feedback circuit: Texas Instruments OPA277; PD: Thorlabs PDA20CS2; Laser: OEwaves HI-Q OE4030; I/Q modulator: Thorlabs LN86S-FC.
    }
    \label{fig:3}
\end{figure}

The schematics of the experimental setup is shown in Fig. \ref{fig:3}. 
A \SI{1560}{nm} laser (OEwaves, Hi-Q OE4030) is fiber coupled to an electro-optic I/Q modulator (Thorlabs, LN86S-FC). 
The \textit{RF}-inputs for the modulator are generated by a circuit that combines primary and auxiliary modulation tones. 
The primary tone with frequency $\Omega_{HF} = 2 \pi \times \SI{3.6}{GHz}$ (chosen for future experimental considerations discussed below) and the auxiliary tone with frequency $\Omega_{LF} = 2 \pi \times \SI{2}{MHz}$ are split  separately with  $90^{\circ}$-hybrid splitters.
Then, the primary and the auxiliary tones are combined into two channels where one channel has an additional phase of $\pi/2$ for both tones.
The input power at the \textit{RF}-ports for the $\Omega_{HF}$-modulation is \SI{23}{dBm} corresponding to a modulation depth of $\beta_{HF} \approx 2.23$ -- defined as $\beta=\pi(V_{mod}/V_{\pi})$, where $V_{\pi}$ is the modulator's half-wave voltage ($V_{\pi} = \SI{4.5}{V}$). 
For the auxiliary modulation, the power is \SI{-6}{dBm} corresponding to a modulation depth of $\beta_{LF} \approx 0.08$, making the weak modulation analysis (Eq. \ref{eq:MZI_out} and Eq. \ref{eq:P_out}) applicable.

Fig. \ref{fig:2}b-c show the spectrum of the output of the I/Q modulator in CS-SSB mode. 
For this measurement, the laser frequency was ramped through manipulation of the laser current and the transmission through a home-built \SI{63}{kHz}-linewidth cavity was detected. 
The optical carrier and the upper $\Omega_{HF}$-SB are suppressed by \SI{-29.3}{dB} and \SI{-30.7}{dB} respectively, relative to the maximized lower $\Omega_{HF}$-SB.
The auxiliary modulation generates a weak CS-SSB with a \SI{-25.5}{dB} peak relative to the maximized $\Omega_{HF}$-SB.

In order to detect the beat signal, \SI{100}{\micro W} of the modulator's output (out of \SI{0.9}{mW}) is split-off with a beam splitter and detected on a photodiode (Thorlabs, PDA20CS2, \SI{11}{MHz} bandwidth). 
The photodiode signal is fed to the demodulation circuit (Fig. \ref{fig:3}), where the three phase bias error signals are obtained by mixing down the input with the appropriate beat frequencies. 
The phase bias errors $\delta_A$ and $\delta_B$ are derived from the \SI{2}{MHz} quadratures. 
Shifting the phase $\theta_2$ of the global \SI{2}{MHz} demodulation tone and a following $90^\circ$-power splitting gives access to the quadratures and enables decoupling of the error signals.
Similarly, the \SI{4}{MHz} demodulation tone is generated by frequency doubling and appropriately shifting the phase $\theta_1$.
Not shown in the schematics are the attenuators and amplifiers (Mini-circuits, MAN-1LN) which are required to ensure the correct input power levels for the shown components.

\begin{figure}[b]
    \centering
    \includegraphics[width=\linewidth]{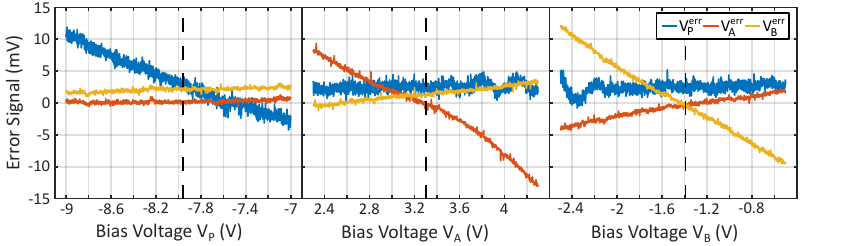}
    \caption{Error signals over phase bias voltage scan. 
    For evaluation of the error signals $V^{err}_{A,B,P}$, the phase bias voltages ($V^{DC}_{A,B,P}$) were scanned through their set points (dashed black lines) and the signals that were generated by the demodulation circuit were measured.
    Left: $V^{err}_{A,B}$ are decoupled from changes in $V^{DC}_{P}$, indicated by absence of a gradient. 
    Offsets can be compensated by the adjusting the set points ($V^{set}_{A,B,P}$) with the feedback circuit.
    Center and right: Scans of $V^{DC}_{A}$ and $V^{DC}_{B}$ show a minor cross talk between $V^{err}_{A}$ and $V^{err}_{B}$, which originates from the fixed imperfect phase ($\sim \SI{90}{\degree}$) between the LO signals used for demodulation. }
    \label{fig:4}
\end{figure}

Fig. \ref{fig:4} shows measurements of the phase bias error signals. 
The phase biases were sequentially modulated around their set points for the CS-SSB operation and the error signals ($V^{err}_{A,B,P}$) generated by the demodulation circuit were recorded with an oscilloscope. 
The plots show that changes in each individual phase bias mainly affect only one of the three error signals. 
The optimization of the demodulation phases $\theta_{1,2}$ are facilitated through such measurements.
A minor cross talk between the two \SI{2}{MHz} quadratures can be observed.
It originates from the fixed imperfect phase between the demodulation tones given by the $90^\circ$-power splitter (Fig. \ref{fig:3}, demodulation circuit).
Nevertheless this does not cause a problem in phase bias locking, since every signal has its own dominant channel.

A home-built, 3-channel, analog feedback circuit simultaneously controls the phase bias voltages ($V^{DC}_{A,B,P}$). 
Since the expected drifts are slow, the bandwidth of the feedback circuit was designed to be around \SI{100}{Hz}.
In order to lock the modulator, the phase bias set points ($V^{set}_{A,B,P}$) are adjusted manually. 
By engaging the lock, the circuit adds the feedback signals to the corresponding set point voltages. 

The error signals in Fig. \ref{fig:4} have small offsets. 
In order to compensate those offsets, the zero-crossing of the error signals are adjustable by adding offsets to the error signals prior to the integrating part of the circuit.
For all mathematical operations to perform feedback we use OPA277 operational amplifiers, and avoid ground loops between connected instruments by using differential amplifiers (AMP03) for all inputs and outputs.

\section{Results}

\subsection{Feedback stability}
\begin{figure}[t]
    \centering
    \includegraphics[width=\linewidth]{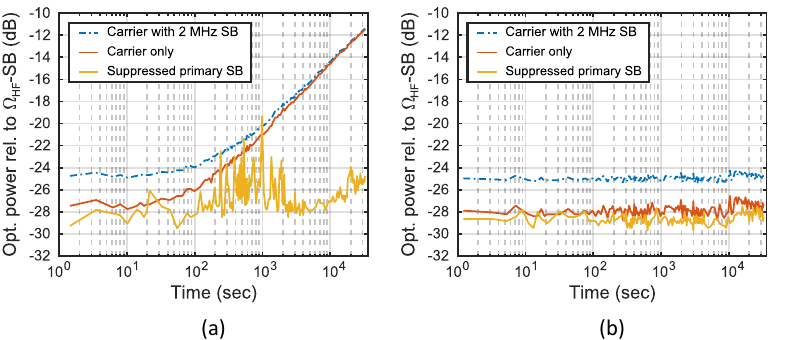}
    \caption{CS-SSB modulation stability over time ($\sim \SI{9}{h}$). 
    The logarithmic time scale demonstrates the short term and long term behaviours of the suppressed primary SB (yellow), the suppressed carrier including the \SI{2}{MHz}-SB (dashed blue), and the inferred carrier only (red). 
    Left: Example trace for the unlocked modulator. 
    The carrier drifts from \SI{-27.4}{dB} to \SI{-11.6}{dB} relative to the unsuppressed $\Omega_{HF}$-SB power and the suppressed SB fluctuates between \SI{-29.3}{dB} and \SI{-19.4}{dB}.
    Right: Trace of the locked modulator.
    Suppressed carrier and SB are stabilized below \SI{-27}{dB} relative to the unsuppressed $\Omega_{HF}$-SBB power.
    }
    \label{fig:5}
\end{figure}

We evaluate the stability of the presented locking method by tracking the powers of the suppressed optical carrier and SB over the course of \SI{9}{h}.
The optical powers are measured with a scanning Fabry-Perot interferometer (Thorlabs, SA200-12B).
Fig. \ref{fig:5}a shows an example for the unlocked modulator in SC-SSB mode. 
The carrier power drifts from \SI{-27.4}{dB} to \SI{-11.6}{dB} while the upper $\Omega_{HF}$-sideband drifts drifts between \SI{-29.3}{dB} and \SI{-19.4}{dB}.
Fig. \ref{fig:5}b shows the measurement of the suppressed optical frequency components for the locked modulator.
The feedback keeps the suppression of the carrier and upper SB below \SI{-27}{dB}. 
The locked modulator was running in laboratory conditions for several weeks without loosing the set point. 
Note that the linewidth (\SI{7.5}{MHz}) of the Fabry-Perot interferometer used for these particular measurements was too large to distinguish the auxiliary SBs from the carrier. 
Therefore, the bare measurements indicate the sum of the powers in the auxiliary SBs and the carrier. 
However, the weak auxiliary modulation with a maximum intensity of \SI{-25.5}{dB} with respect to the unsuppressed $\Omega_{HF}$-sideband (Fig. \ref{fig:2}c) allows us to calculate the power in the carrier alone. 
In principle the auxiliary sidebands can be made much smaller, however this was not pursued since the residual sideband levels are sufficient for our intended applications.
As a separate note, we observed that the drifts in the suppression levels do not originate directly from temperature drifts, since deliberate temperature changes larger than ambient fluctuations did not cause similar effects. 

\section{Discussion and Conclusion}

In this work, we designed and implemented a phase bias stabilization method for an electro-optic I/Q modulator. 
A weak auxiliary modulation was applied to generate optical \SI{2}{MHz}-sidebands besides the primary $\Omega_{HF}$-sidebands which can range up to \SI{10}{GHz}.
We showed that all three involved phase bias errors can be deduced from the beat signals between these auxiliary sidebands and the optical carrier. 
The configuration of the demodulation circuit was discussed in detail and it was shown that the generated error signals are decoupled from each other by tuning the relevant demodulation phases.
Furthermore, we described a home-built, all-analog feedback circuit, that enables locking of the phases at their desired values. 
Long-term stabilization of the modulator in CS-SSB mode was demonstrated. 
The optical carrier and the suppressed sideband were kept below \SI{-27}{dB} relative to the unsuppressed $\Omega_{HF}$-sideband.
These values were also the best obtainable in absence of any feedback in the current device.

The demonstrated setup is intended for use in generation of frequency-offset counter-propagating laser tones that will derive two-photon Raman transitions in an atom interferometer experiment to realize atomic beam splitters and mirrors \cite{Jaffe2018}. 
In our experiment, $^{87}Rb$ atoms will be dipole-trapped by an optical cavity mode with an ultra-narrow linewidth \SI{1560}{nm} laser.
This low noise laser will also serve to generate the Raman beams after frequency doubling to \SI{780}{nm} to near-resonantly address the atoms.
One beam will be generated by directly frequency doubling, and the other one by first shifting the frequency with the I/Q modulator by \SI{3.417}{GHz} with CS-SSB generation. 
After frequency doubling and amplifying, the frequency offset between the two Raman beams will match exactly the hyperfine clock transition of $^{87}Rb$.  
Stabilization of the modulator at high suppression levels is crucial for the efficiency of the operations performed on the atoms due to the adverse effects of parasitic tones on the energy level shifts \cite{Carraz2012}.

Compared to other locking schemes \cite{Fabbri2013,Templier2021} that rely on dithering the phase bias control voltages in the low \SI{}{kHz}-range, the presented method is less prone to various technical ($1/f$) noise sources, and hence could in principle operate with smaller modulation levels.
Besides that, these methods generate \SI{}{kHz}-sidebands that are difficult to filter from the optical carrier. 
Hence, our method might be advantageous for high-precision applications, where the carrier is further utilized, e.g. the SSB configuration.
Lastly, the analog nature of the feedback eliminates unwanted noise due to digitization and spurious digital noise.

Our feedback scheme is not limited to CS-SSB modulation of an I/Q modulator. 
The tunable demodulation phases in combination with the adjustable error signal offsets allow to optimize the error signals for any intended set point. 
In principle, the developed method can be applied to stabilize amplitude modulators and it might be possible to stabilize modulators with more complex structures by advanced configurations for the auxiliary modulation and phase bias error demodulation.

\section*{Acknowledgments}
This work was supported by Institute of Science and Technology (ISTA) Austria. We thank Jakob Vorlaufer for technical contributions and Vyacheslav Li and Sofia Agafonova for comments on the manuscript. 

\section*{Disclosures}
The authors declare no conflict of interest.

\section*{Data availability}
Data underlying the results presented in this paper are not publicly available at this time but might be obtained or accessed in the future.

\bibliography{Analog_Stabilization_of_an_Electro-Optic_IQ_Modulator}

\end{document}